\documentstyle[aps, preprint]{revtex}

\begin{document}
\title{Role of gauge invariance in $B\rightarrow V\gamma $ radiative weak decays}
\author{Riazuddin}
\address{National Center for Physics, Quaid-e-Azam University, Islamabad 45320,\\
Pakistan}
\maketitle

\begin{abstract}
The role of gauge invariance in calculating $B\rightarrow V\gamma $
radiative weak decays is clarified. It is shown that the gauge invariance,
particularly in the context of vector meson dominance, severely restricts
the contributions mediated by the usual weak non-leptonic Hamiltonian
dominated by $u$ and $c$ quarks with one photon attachment. We also
discussed contributions which are not restricted to vector dominance.

\end{abstract}

The purpose of this paper is to clarify the role of gauge invariance in
calculating $B\rightarrow V\gamma $ radiative weak decays in a general way
by usual Ward identities. In particular it is shown that the gauge
invariance in the context of vector meson dominance and factorization
severely restricts the contributions dominated by $u$ and $c$ quarks. Such
contributions are controlled by the effective weak interaction Hamiltonian 
\cite{baver} 
\begin{eqnarray}
H_{eff}^{w} &=&\frac{G_{F}}{\sqrt{2}}\left\{ \lambda _{u}^{(q)}\left[
a_{1}\left( \bar{u}b\right) _{V-A}\left( \bar{q}u\right) _{V-A}+a_{2}\left( 
\bar{q}b\right) _{V-A}\left( \bar{u}u\right) _{V-A}\right] \right.  \nonumber
\label{01} \\
&&+\left. \lambda _{c}^{(q)}\left[ a_{1}\left( \bar{c}b\right) _{V-A}\left( 
\bar{q}c\right) _{V-A}+a_{2}\left( \bar{q}b\right) _{V-A}\left( \bar{c}%
c\right) _{V-A}\right] \right\}  \label{01}
\end{eqnarray}
where $q=d$ or $s$, $\left( \bar{u}b\right) _{V-A}=\bar{u}\gamma _{\mu
}\left( 1-\gamma _{5}\right) b$ etc., $a_{1}=c_{1}\left( m_{b}\right)
+c_{2}\left( m_{b}\right) \frac{1}{N_{c}}=1.02,$ $a_{2}=c_{2}\left(
m_{b}\right) +c_{1}\left( m_{b}\right) \frac{1}{N_{c}}=0.17$ \cite{buchalla}%
, $\lambda _{u}^{(q)}=V_{ub}^{*}V_{uq}$, $\lambda
_{c}^{(q)}=V_{cb}^{*}V_{cq} $.

We will consider $B\rightarrow \rho \gamma $ as the prototype for our
discussion for which we can write the matrix elements to first order in
electromagnetism $\left( e>0,-e\text{ charge of }B^{-}\right) $ and to all
orders in the strong coupling: 
\begin{eqnarray}
\left\langle \gamma \left( \epsilon ,q\right) \rho \left( \eta ,k\right)
\left| H_{eff}^w\right| \bar{B}\left( p\right) \right\rangle &=&ie\epsilon
^{\lambda *}\left( q\right) \int d^4xe^{iq\cdot x}\left\langle \rho \left(
\eta ,k\right) \left| T\left( j_\lambda ^{em}\left( x\right)
H_{eff}^w\right) \right| \bar{B}\left( p\right) \right\rangle  \nonumber \\
&\equiv &ie\epsilon ^{\lambda *}\left( q\right) M_\lambda  \label{02}
\end{eqnarray}

The conservation of electromagnetic current gives the following Ward
identity 
\begin{eqnarray}
-iq^{\lambda }M_{\lambda } &=&\int d^{3}xe^{i{\bf q}\cdot {\bf x}%
}\left\langle \rho \left( \eta ,k\right) \left| \left[ j_{0}^{em}\left( {\bf %
x},0\right) ,H_{eff}^{w}\right] \right| \bar{B}\left( p\right) \right\rangle
\nonumber \\
&=&\left\langle \rho \left( \eta ,k\right) \left| \left[ Q^{em}\left( {\bf x}%
,0\right) ,H_{eff}^{w}\right] \right| \bar{B}\left( p\right) \right\rangle 
\nonumber \\
&=&0  \label{03}
\end{eqnarray}
since $\bar{B}$ and $\rho $ have the same charge 0 or $-1$. Now the most
general parameterization of $M_{\lambda }$ is 
\begin{equation}
M_{\lambda }=A\eta _{\lambda }^{*}+q\cdot \eta ^{*}\left( Bp_{\lambda
}+Cq_{\lambda }\right) +i\epsilon _{\lambda \mu \alpha \beta }\eta ^{\mu
*}p^{\alpha }q^{\beta }D  \label{04}
\end{equation}
The Ward identity (\ref{03}) then gives 
\begin{equation}
A+Bp\cdot q+Cq^{2}=0  \label{05}
\end{equation}
Thus we can write $M_{\lambda }$ as 
\begin{equation}
M_{\lambda }=B\left( -p\cdot q\eta _{\lambda }^{*}+q\cdot \eta
^{*}p_{\lambda }\right) +C\left( q\cdot \eta ^{*}q_{\lambda }-q^{2}\eta
_{\lambda }^{*}\right) +i\epsilon _{\lambda \mu \alpha \beta }\eta ^{\mu
*}p^{\alpha }q^{\beta }D  \label{06a}
\end{equation}
For the real photon on the mass shell 
\begin{equation}
\left\langle \gamma \left( \epsilon ,q\right) \rho \left( \eta ,k\right)
\left| H_{eff}^{w}\right| \bar{B}\left( p\right) \right\rangle =ie\left\{
\left( -p\cdot q\epsilon ^{*}\cdot \eta ^{*}+p\cdot \epsilon ^{*}q\cdot \eta
^{*}\right) B+i\epsilon _{\lambda \mu \alpha \beta }\epsilon ^{\lambda
*}\eta ^{\mu *}p^{\alpha }q^{\beta }D\right\}  \label{06}
\end{equation}

The question of gauge invariance in calculating the photon coupling to
hadrons has been discussed by Kroll-Lee-Zumino \cite{kroll} and Sakurai \cite
{Sakurai}, particularly the compatibility of the vector meson dominance(VMD)
with the gauge invariance. According to their scheme there are two
equivalent ways of formulating photon coupling to hadrons: (i) the
explicitly gauge invariant formulation involving direct gauge-invariant
coupling to quarks and no VMD coupling $V.A^{\left( \gamma \right) }$ ($%
A^{\left( \gamma \right) }$ is the electromagnetic field). An example of
this is the $t$-quark dominated contribution: 
\begin{equation}
-\frac{2G_{F}}{\sqrt{2}}\frac{e}{16\pi ^{2}}\lambda _{t}^{\left( q\right)
}C_{7}\left( \mu \right) m_{b}\bar{q}\sigma ^{\mu \nu }\left( 1+\gamma
_{5}\right) bF_{\mu \nu }
\end{equation}
which dominates the radiative $B\rightarrow V\gamma $ decays. This is not
possible for any of the operators in $(1)$ e.g. $\bar{q}\gamma _{{}}^{\mu
}\left( 1-\gamma _{5}\right) bA_{\mu }^{\left( \gamma \right) }$ is not
gauge invariant at quark level \cite{deshpande} for $b\rightarrow q\gamma $.
(ii) the standard VMD formulation, which states that the coupling of photon
to hadrons may be obtained by first calculating the vector-meson coupling
and then performing the substitution 
\begin{equation}
V_{\mu }=\frac{e}{g_{V}}A_{\mu }^{\left( \gamma \right) },\text{ }g_{V}=%
\frac{m_{V}^{2}}{f_{V}},\text{ e.g. }V=\rho
\end{equation}
Here the gauge invariance is not explicit. We write the Hamiltonian (\ref{01}%
) generically as 
\begin{equation}
\frac{G_{F}}{\sqrt{2}}\left[ J_{1}\cdot J_{2}+J_{1}^{\prime }\cdot
J_{2}^{\prime }\right]
\end{equation}
where e.g. $J_{1}\cdot J_{2}$ refers to the first and third terms while $%
J_{1}^{\prime }\cdot J_{2}^{\prime }$ refers to the second and fourth terms
in Eq. (\ref{01}). Through the use of field current--identity [e.g. $\bar{c}%
\gamma ^{\mu }c\rightarrow f_{\psi }\psi ^{\mu }$] one obtains \cite{sarraga}
either

\begin{enumerate}
\item[a)]  : 
\begin{equation}
\frac{G_{F}}{\sqrt{2}}J_{1}.V_{2}
\end{equation}
[$V_{2}$ being the vector part of $J_{2}$] with vanishing direct
photon-current coupling and $V_{2}.A^{\left( \gamma \right) }$ VMD coupling
or

\item[b)]  : 
\begin{equation}
\frac{G_{F}}{\sqrt{2}}\left[ J_{1}^{\prime }.V_{2}^{\prime }-\frac{em_{V}^{2}%
}{g_{V}}J_{1}^{\prime }.A^{\left( \gamma \right) }\right]
\end{equation}
where the second contribution cancels the $V_{2}^{\prime }.$ $A^{\left(
\gamma \right) }$ VMD coupling. Thus in either case the gauge invariance in
VMD gives null result.
\end{enumerate}

We now apply the above general considerations to $B\rightarrow V\gamma $,
mediated by the weak Hamiltonian (1). The generation of direct coupling of $%
\bar{B}$ with the photon can be seen as follows: Using the field current
identity $\bar{c}\gamma _{\lambda }c\rightarrow f_{\psi }\psi _{\lambda }$, $%
\left\langle 0\left| \bar{c}\gamma _{\lambda }c\right| \psi \right\rangle
=f_{\psi }\epsilon _{\lambda }^{\psi }$ and then introducing
electromagnetism $\psi _{\lambda }\rightarrow \psi _{\lambda }+\frac{2}{3}%
\frac{ef_{\psi }}{m_{\psi }^{2}}A_{\lambda }^{\gamma }$ and similarly for $%
\bar{u}\gamma _{\lambda }u$ term through $\rho ^{0}\equiv \frac{1}{\sqrt{2}}%
\left( \bar{u}u-\bar{d}d\right) $ and $\omega ^{0}\equiv \frac{1}{\sqrt{2}}%
\left( \bar{u}u+\bar{d}d\right) $ [$\left\langle 0\left| \bar{u}u\right|
\rho ^{0}\right\rangle \equiv f_{\rho ^{0}}=\left\langle 0\left| \bar{u}%
u\right| \omega \right\rangle \equiv f_{\omega }$] mesons, we see that the
Hamiltonian $H_{em}=eJ_{\lambda }^{em}A_{\lambda }^{\gamma }$ in the
presence of weak interaction goes over to 
\begin{equation}
H_{em}=eJ_{\lambda }^{em}A^{\gamma \lambda }+e\frac{G_{F}}{\sqrt{2}}%
a_{2}\left\{ \left[ \frac{f_{\rho ^{0}}^{2}}{m_{\rho }^{2}}+\frac{f_{\omega
}^{2}}{m_{\omega }^{2}}\right] \left( \bar{d}\gamma _{\lambda }\left(
1-\gamma _{5}\right) b\right) A^{\gamma \lambda }+\frac{2}{3}\frac{f_{\psi
}^{2}}{m_{\psi }^{2}}\left( \bar{d}\gamma _{\lambda }\left( 1-\gamma
_{5}\right) b\right) A^{\gamma \lambda }\right\}  \label{07}
\end{equation}
Thus the long distance contribution through $\psi $ has two parts: one
arising from the third (direct coupling) term in equation (\ref{07}) and the
other through $V.A^{\left( \gamma \right) }$ coupling: $\left\langle \gamma
\mid \psi \right\rangle \left\langle \psi \mid H_{w}\mid \bar{B}\rho
\right\rangle $which in the factorization approximation gives 
\begin{equation}
\frac{2}{3}e\frac{G_{F}}{\sqrt{2}}a_{2}\left\{ f_{\psi }\left\langle \rho
\left| \bar{d}\gamma ^{\mu }\left( 1-\gamma _{5}\right) b\right| \bar{B}%
\right\rangle \left( g_{\mu \lambda }-\frac{q_{\mu }q_{\lambda }}{m^{2}}%
\right) \frac{f_{\psi }}{q^{2}-m_{\psi }^{2}}+\left\langle \rho \left| \bar{d%
}\gamma _{\lambda }\left( 1-\gamma _{5}\right) b\right| \bar{B}\right\rangle 
\frac{f_{\psi }^{2}}{m_{\psi }^{2}}\right\} \epsilon ^{\lambda ^{*}}
\label{08}
\end{equation}
This vanishes for the real photon; the contribution from the direct term in
equation (\ref{08}) exactly cancels the first contribution. Similarly the
contributions through $\rho ^{0}$and $\omega $ mesons also vanish \cite
{sarraga}. Thus the contributions labelled as $P_{c}$ and $P_{u}$ in ref. 
\cite{grinstein} are exactly zero in the factorization approximation used
there. In fact the authors of this reference and that of \cite{cheng} have
to force gauge invariance by requiring a particular relation between $%
A_{1}\left( 0\right) $ and $A_{2}\left( 0\right) $ which appears in the
parametrization of $\left\langle \rho \left| \bar{d}\gamma _{\mu }\left(
1-\gamma _{5}\right) b\right| \bar{B}\right\rangle $, and which is not a
consequence of any symmetry principle.

Let us now consider the first part of the Hamiltonian (\ref{01}), which for
the process $B^{-}\rightarrow \rho ^{-}\gamma $, in the factorization ansatz
gives 
\begin{eqnarray}
\left\langle \rho ^{-}\gamma \left| H_{eff}^{w}\right| B^{-}\right\rangle &=&%
\frac{G_{F}}{\sqrt{2}}\lambda _{u}^{d}a_{1}\left\{ -if_{B}p^{\mu
}\left\langle \rho ^{-}\gamma \left| \bar{d}\gamma _{\mu }\left( 1-\gamma
_{5}\right) u\right| 0\right\rangle +f_{\rho ^{-}}\eta ^{\mu }\left\langle
\gamma \left| \bar{u}\gamma _{\mu }\left( 1-\gamma _{5}\right) b\right|
B^{-}\right\rangle \right\}  \nonumber \\
&=&\frac{G_{F}}{\sqrt{2}}\lambda _{u}^{d}a_{1}\left\{ -if_{B}p^{\mu }\left(
-ie\right) \epsilon ^{\lambda *}\left( q\right) N_{1\lambda \mu }+f_{\rho
^{-}}\eta ^{\mu *}\epsilon ^{\lambda *}\left( q\right) \left( -ie\right)
N_{2\lambda \mu }\right\}  \label{09}
\end{eqnarray}
where with $J_{1\mu }=\bar{d}\gamma _{\mu }\left( 1-\gamma _{5}\right) u$, $%
J_{2\mu }=\bar{u}\gamma _{\mu }\left( 1-\gamma _{5}\right) b$%
\begin{eqnarray}
N_{1\lambda \mu } &=&\int e^{iq\cdot x}\left\langle \rho ^{-}\left| T\left(
j_{\lambda }^{em}\left( x\right) J_{1\mu }\left( 0\right) \right) \right|
0\right\rangle d^{4}x  \nonumber \\
&=&\int e^{-ip\cdot x}\left\langle \rho ^{-}\left| T\left( j_{\lambda
}^{em}\left( 0\right) J_{1\mu }\left( x\right) \right) \right|
0\right\rangle d^{4}x  \label{10} \\
N_{2\lambda \mu } &=&\int e^{iq\cdot x}\left\langle 0\left| T\left(
j_{\lambda }^{em}\left( x\right) J_{2\mu }\left( 0\right) \right) \right|
B^{-}\right\rangle d^{4}x  \label{11}
\end{eqnarray}
Since $J_{1\mu }\left( x\right) $ is conserved in the chiral limit, the Ward
identities give 
\begin{eqnarray}
ip^{\mu }N_{1\lambda \mu } &=&-\left\langle \rho ^{-}\left| \left[
j_{\lambda }^{em},I_{+}\right] \right| 0\right\rangle  \nonumber \\
&=&\sqrt{2}\left\langle \rho ^{0}\left| j_{\lambda }^{em}\right|
0\right\rangle  \nonumber \\
&=&\sqrt{2}f_{\rho ^{0}}\eta _{\lambda }^{*}=f_{\rho ^{-}}\eta _{\lambda
}^{*}  \label{12} \\
-iq^{\lambda }N_{2\lambda \mu } &=&\left\langle 0\left| \left[
Q^{em},J_{2\mu }\right] \right| B^{-}\right\rangle  \nonumber \\
&=&\left\langle 0\left| J_{2\mu }\right| B^{-}\right\rangle  \nonumber \\
&=&-if_{B}p_{\mu }  \label{13}
\end{eqnarray}
Thus we write 
\begin{equation}
N_{2\lambda \mu }=f_{B}\frac{\left( 2p-q\right) _{\lambda }p_{\mu }}{2p\cdot
q-q^{2}}+\tilde{N}_{2\lambda \mu }  \label{14}
\end{equation}
where $q^{\lambda }\tilde{N}_{2\lambda \mu }=0.$ Hence we obtain $(q^{2}=0)$%
\begin{equation}
\left\langle \rho ^{-}\gamma \left| H_{eff}^{w}\right| B^{-}\right\rangle =ie%
\frac{G_{F}}{\sqrt{2}}\lambda _{u}^{d}a_{1}\left\{ f_{B}f_{\rho ^{-}}\eta
^{*}\cdot \epsilon ^{*}-f_{B}f_{\rho ^{-}}\frac{p\cdot \epsilon ^{*}p\cdot
\eta ^{*}}{p\cdot q}-f_{\rho ^{-}}\eta ^{\mu *}\epsilon ^{\lambda *}\tilde{N}%
_{2\lambda \mu }\right\}  \label{15}
\end{equation}
Now the most general parametrization of $\tilde{N}_{2\lambda \mu }$ is 
\begin{equation}
\tilde{N}_{2\lambda \mu }=g_{\lambda \mu }f_{1}+p_{\lambda }p_{\mu
}f_{2}+q_{\lambda }q_{\mu }f_{3}+q_{\lambda }p_{\mu }f_{4}+p_{\lambda
}q_{\mu }f_{5}-i\epsilon _{\lambda \mu \alpha \beta }p^{\alpha }q^{\beta
}f_{V}  \label{16}
\end{equation}
so that $iq^{\lambda }\tilde{N}_{2\lambda \mu }=0$ give 
\begin{equation}
f_{1}+q^{2}f_{3}+p\cdot qf_{5}=0,\text{ }q\cdot pf_{2}+q^{2}f_{4}=0
\label{17}
\end{equation}
Thus we can write 
\begin{equation}
\tilde{N}_{2\lambda \mu }=\left( -p\cdot qg_{\lambda \mu }+p_{\lambda
}q_{\mu }\right) f_{5}+\left( q_{\lambda }q_{\mu }-q^{2}g_{\lambda \mu
}\right) f_{3}+\left( q_{\lambda }p_{\mu }-\frac{q^{2}}{p\cdot q}p_{\lambda
}p_{\mu }\right) f_{4}-i\epsilon _{\lambda \mu \alpha \beta }p^{\alpha
}q^{\beta }f_{V}  \label{18}
\end{equation}
Hence using Eq. (\ref{15}) $\left[ p\cdot \eta ^{*}=q\cdot \eta ^{*}\right] $
\cite{grinstein} 
\begin{equation}
\left\langle \rho ^{-}\gamma \left| H_{eff}^{w}\right| B^{-}\right\rangle =ie%
\frac{G_{F}}{\sqrt{2}}\lambda _{u}^{d}a_{1}f_{\rho ^{-}}\eta ^{\mu
*}\epsilon ^{\lambda *}\left\{ \left( p\cdot qg_{\lambda \mu }-p_{\lambda
}q_{\mu }\right) \left( \frac{f_{B}}{p\cdot q}+f_{5}\right) +i\epsilon
_{\lambda \mu \alpha \beta }p^{\alpha }q^{\beta }f_{V}\right\}  \label{19}
\end{equation}
which is manifestly gauge invariant.

The next question is the evaluation of the form factors $f_{5}$ and $f_{V}$.
One way of doing this is through vector ($\rho ^{0}$) dominance in $t=q^{2}$
channel, which is a good example to show how non-gauge invariant terms are
cancelled out in this case. Thus 
\begin{equation}
\eta ^{\mu *}\tilde{N}_{2\lambda \mu }=\left( g_{\nu \lambda }-\frac{q_{\nu
}q_{\lambda }}{q^{2}}\right) \eta _{\mu }^{*}M^{\mu \nu }\frac{f_{\rho }}{%
q^{2}-m_{\rho }^{2}}  \label{20}
\end{equation}
where 
\begin{eqnarray}
\epsilon _{\lambda }^{*}M^{\lambda \mu } &=&i\left\langle \rho ^{0}\left(
q\right) \left| J_{2}^{\mu }\right| B^{-}\left( p\right) \right\rangle 
\nonumber \\
&=&i\epsilon _{\lambda }^{*}\left\{ -\epsilon ^{\lambda \mu \alpha \beta
}p_{\alpha }q_{\beta }\frac{2V\left( k^{2}\right) }{m_{B}+m_{\rho }}-i\left[
\left( g^{\lambda \mu }-\frac{k^{\lambda }k^{\mu }}{k^{2}}\right) \left(
m_{B}+m_{\rho }\right) A_{1}\left( k^{2}\right) \right] \right.  \nonumber \\
&&\left. -\left[ \left( p+q\right) ^{\mu }-\frac{m_{B}^{2}-m_{\rho }^{2}}{%
k^{2}}k^{\mu }\right] k^{\lambda }\frac{A_{2}\left( k^{2}\right) }{%
m_{B}+m_{\rho }}+k^{\lambda }k^{\mu }\frac{2m_{\rho }}{k^{2}}A_{0}\left(
k^{2}\right) \right\}  \label{21}
\end{eqnarray}
Then using $k\cdot \eta ^{*}=0$ [$\hat{F}$'s are certain combinations of $A$%
's, see below] 
\begin{eqnarray}
\eta ^{\mu *}\tilde{N}_{2\lambda \mu }\left( \rho ^{0}\right) &=&\frac{%
f_{\rho }}{q^{2}-m_{\rho }^{2}}\left\{ -i\epsilon _{\lambda \mu \alpha \beta
}\eta ^{\mu *}p^{\alpha }q^{\beta }\frac{2V}{m_{B}+m_{\rho }}+\left( -p\cdot
q\eta _{\lambda }^{*}+p_{\lambda }q\cdot \eta ^{*}\right) \hat{F}_{5}+\hat{F}%
_{3}\left( -q^{2}\eta _{\lambda }^{*}+q_{\lambda }q\cdot \eta ^{*}\right)
\right\}  \nonumber \\
&&+\frac{f_{\rho }}{m_{\rho }^{2}}\left[ -\eta _{\lambda }^{*}\left( \hat{F}%
_{1}-\left( -p\cdot q+q^{2}\right) \hat{F}_{5}\right) \right]  \label{22}
\end{eqnarray}
where the non-gauge invariant terms in square bracket combine with the
direct term to preserve the gauge invariance. The direct term is given by 
\begin{equation}
\left\langle \gamma \left| J_{2\mu }\right| B^{-}\right\rangle =\frac{%
f_{\rho }}{m_{\rho }^{2}}\epsilon ^{\lambda *}\left\langle \rho ^{0}\left|
J_{2\mu }\right| B^{-}\right\rangle =-i\epsilon ^{\lambda *}\tilde{N}%
_{2\lambda \mu }^{\text{direct}}  \label{23}
\end{equation}
This gives 
\begin{equation}
\eta ^{\mu *}\tilde{N}_{2\lambda \mu }^{\text{direct}}=\frac{f_{\rho }}{%
m_{\rho }^{2}}\left\{ -i\epsilon _{\lambda \mu \alpha \beta }\eta ^{\mu
*}p^{\alpha }q^{\beta }\frac{2V}{m_{B}+m_{\rho }}+\left[ \eta _{\lambda }^{*}%
\hat{F}_{1}+\hat{F}_{5}q\cdot \eta ^{*}\left( p-q\right) _{\lambda }\right]
\right\}  \label{24}
\end{equation}
Combined with (\ref{22}), we finally obtain fully gauge invariant
expression. For real photons for which only non-zero covariants are $%
i\epsilon _{\lambda \mu \alpha \beta }\eta ^{\mu *}p^{\alpha }q^{\beta }$
and $\left( -p\cdot q\eta _{\lambda }^{*}+p_{\lambda }q\cdot \eta
^{*}\right) ,$%
\begin{eqnarray}
\eta ^{\mu *}\tilde{N}_{2\lambda \mu } &=&f_{\rho }\left( \left. \frac{1}{%
q^{2}-m_{\rho }^{2}}\right| _{q^{2}=0}+\frac{1}{m_{\rho }^{2}}\right) 
\nonumber \\
&&\times \left\{ -i\epsilon _{\lambda \mu \alpha \beta }\eta ^{\mu
*}p^{\alpha }q^{\beta }\frac{2V}{m_{B}+m_{\rho }}+\left( -p\cdot q\eta
_{\lambda }^{*}+p_{\lambda }q\cdot \eta ^{*}\right) \hat{F}_{5}\right\}
\label{25}
\end{eqnarray}
giving $f_{V}$ and $f_{5}$ to be zero. Thus the only surviving term in Eq. (%
\ref{19}) is 
\begin{equation}
\left\langle \tilde{\rho}\gamma \left| H_{eff}^{w}\right| B^{-}\right\rangle
=ie\frac{G_{F}}{\sqrt{2}}\lambda _{u}^{d}a_{1}\eta ^{\mu *}\epsilon
^{\lambda ^{*}}\left( p\cdot qg_{\lambda \mu }-p_{\lambda }q_{\mu }\right) 
\frac{f_{B}f_{\rho ^{-}}}{\left( p\cdot q\right) }  \label{26}
\end{equation}
This also implies that in this approximation 
\begin{equation}
\left\langle \rho ^{0}\gamma \left| H_{eff}^{w}\right| \bar{B}%
^{0}\right\rangle =0  \label{27}
\end{equation}
We emphasize that these results have been obtained within the framework of
VMD and the factorization ansatz.

There are, however other contributions to $f_{V}$ and $f_{5}$, for example, $%
B^{*}(1^{-})$, $B_{A}^{*}(1^{+})$ etc. poles in the channel $s=k^{2}$ or
even from the continuum (e.g. quark triangle). Instead of calculating these
contributions explicitly, we notice that the form factor $F_{1}$ appearing
in $\left\langle \rho \left| \bar{d}\sigma _{\mu \nu }q^{\nu }b\right|
B^{-}\right\rangle $ and those in $\left\langle \gamma \left| \bar{u}\gamma
_{\mu }\left( 1-\gamma _{5}\right) b\right| B^{-}\right\rangle $ are related
through the universal function $\xi _{\bot }$ \cite{charles} 
\begin{eqnarray}
&&\left. F_{1}\left( q^{2}=\left( p_{B}-p_{\rho }\right) ^{2}=0\right)
\approx \xi _{\bot }\left( 0\right) \right.  \label{28} \\
&&\left. f_{V}=f_{5}\text{ }\left( \equiv f_{A}\right) =\frac{f_{\rho ^{-}}}{%
m_{\rho }^{2}}\frac{2}{3}\left( 2\xi _{\bot }\left( 0\right) \right) \frac{1%
}{m_{B}}\right.  \label{29}
\end{eqnarray}
where $\xi _{\bot }\left( q^{2}\right) $ scales \cite{charles} as $%
m_{B}^{-3/2}/\left( 1-q^{2}/m_{B}^{2}\right) $ near $q^{2}=0$. Various
estimates \cite{Ali} of $\xi _{\bot }\left( 0\right) $ give it to be $0.20$.
In Eq. $\left( 34\right) $ [and in subsequent equations] $f_{V}$ and $f_{5}$
are evaluated at $\left( p_{B}-p_{\gamma }\right) ^{2}=0$ which corresponds
to $E_{\gamma }=m_{B}/2.$ With $f_{\rho ^{-}}/m_{\rho }\simeq 205$MeV, we
obtain 
\begin{equation}
f_{V}=f_{A}=0.013\text{GeV}^{-1}  \label{30}
\end{equation}
which is of the same order as $\left[ f_{B}\simeq 180\text{MeV}\right] $ $%
f_{B}/p\cdot q=\frac{2f_{B}}{m_{B}}\frac{1}{m_{B}}=0.013$GeV$^{-1}$ [the
first term in Eq. (\ref{19})] and is much smaller than the estimate in \cite
{grinstein} $\left[ E_{\gamma }=m_{B}/2\right] $%
\begin{equation}
f_{V}=\frac{f_{B}}{2m_{B}}\frac{m_{B}}{E_{\gamma }}\frac{2}{3}R\simeq \frac{%
f_{B}}{m_{B}}\frac{2}{3}R\simeq 0.057\text{GeV}^{-1}  \label{31}
\end{equation}
with $R\simeq 2.5$GeV$^{-1}$. Note that $\left( m_{B}f_{V}\right) $ in Eq.(%
\ref{31}) does not satisfy the scaling law $m_{B}^{-3/2}$ and this seems to
be the reason for its enhanced value.

In conclusion we have clarified the role of gauge invariance by using the
Ward identities for contributions involving the Hamiltonian (\ref{01}) in
the factorization ansatz, particularly within the framework of the VMD. The
general case, going beyond $\rho ^{0}$ meason dominance in $t$-channel is
also considered. The contributions for the vector and axial vector form
factor for $B\rightarrow \rho ^{-}\gamma $ are found to be much smaller than
the previous estimate.

The author would like to thank Professor M. A. Virasoro for hospitality at
the {\bf Abdus Salam} International Center for Theoretical Physics, Trieste,
Italy, where most of this work was done.

{\bf Note Added}: After the paper was submitted, I was informed about a
paper \cite{melikhov} in which it has been shown that using gauge invariance
alone, the factorizeable c\={c} contribution in the radiative $B\rightarrow
K^{*}\gamma $ decay vanishes.This reinforces our conclusion reached after
Eq. (\ref{08}) within the framework of VMD, which in fact has been used by
many authors in calculating the photon--coupling to hadrons and where the
role of gauge invariance is not properly treated.

Another recent preprint \cite{Wyler} has also been brought to our attention
where Ward identities have been employed and where although the gauge
invariance in VMD is not discussed but general conclusions seem to agree.

\newpage

\end{document}